\documentclass{ieeeaccess}
\usepackage{cite}
\usepackage{amsmath,amssymb,amsfonts}
\usepackage{algorithmic}
\usepackage{graphicx}
\usepackage{textcomp}

\begin{document}
\history{Date of publication xxxx 00, 0000, date of current version xxxx 00, 0000.}
\doi{10.1109/ACCESS.2017.DOI}

\title{ATP-Net: An Attention-based Ternary Projection Network For Compressed Sensing}
\author{\uppercase{First A. Guanxiong Nie}\authorrefmark{1}, 
\uppercase{Second B. Yajian Zhou\authorrefmark{2}}.}
\address[1]{China Construction Bank Hebei Branch, Hebei Shijiahzuang.050000, China}
\address[2]{Beijing University of Posts and Telecommunications, Beijing, 100876, China}

\markboth
{Author \headeretal: Preparation of Papers for IEEE TRANSACTIONS and JOURNALS}
{Author \headeretal: Preparation of Papers for IEEE TRANSACTIONS and JOURNALS}

\corresp{Corresponding author: First A. Guanxiong Nie (e-mail: niethe10@gmail.com).}

\begin{abstract}
Compressed Sensing (CS) theory simultaneously realizes the signal sampling and compression process, and can use fewer observations to achieve accurate signal recovery, providing a solution for better and faster transmission of massive data. In this paper, a ternary sampling matrix-based method with attention mechanism is proposed with the purpose to solve the problem that the CS sampling matrices in most cases are random matrices, which are irrelative to the sampled signal and need a large storage space. The proposed method consists of three components, i.e., ternary sampling, initial reconstruction and deep reconstruction, with the emphasis on the ternary sampling. The main idea of the ternary method (-1, 0, +1) is to introduce the attention mechanism to evaluate the importance of parameters at the sampling layer after the sampling matrix is binarized (-1, +1), followed by pruning weight of parameters, whose importance is below a predefined threshold, to achieve ternarization. Furthermore, a compressed sensing algorithm especially for image reconstruction is implemented, on the basis of the ternary sampling matrix, which is called ATP-Net, i.e., Attention-based ternary projection network. Experimental results show that the quality of image reconstruction by means of ATP-Net maintains a satisfactory level with the employment of the ternary sampling matrix, i.e., the average PSNR on Set11 is 30.4 when the sampling rate is 0.25, approximately 6\% improvement compared with that of DR2-Net.
\end{abstract}

\begin{keywords}
Compressed sensing,
Deep learning,
Image reconstruction,
Self-Attention,
Ternary
\end{keywords}

\titlepgskip=-15pt

\maketitle

\section{Introduction}
\label{sec:introduction}
\PARstart{I}{mage} sampling systems have to satisfy the classical Nyquist-Shannon sampling theorem with a sampling rate no less than twice the bandwidth of the signal, in order to guarantee any original continuous signal can be completely reconstructed from its samples. However, samples are usually redundant and certain compression operation, which is conventionally computationally complex, is indispensable for removal of redundancy. In the context of some band-restricted applications, e.g., massive data storage, wireless transmission, among others, limitation of the Nyquist-Shannon theorem becomes a major problem.

The emerging technology of compressive sensing (CS) \cite{IEEEexample:donoho2006compressed} \cite{tsaig2006extensions} \cite{candes2006compressive} gives rise to a new paradigm to capture and represent compressible signals, especially for image acquisition and reconstruction, at a rate significantly below the Nyquist rate, i.e., CS theory enables a signal to be recovered from many fewer measurements than that suggested by the Nyquist-Shannon sampling theorem when the signal is sparse in some domain. CS characterizes in implementing the sampling and compression processes jointly, by employs nonadaptive linear projections to preserve the structure of the signal, which is then reconstructed from these projections using an optimization process. In this way, CS is capable of breaking through the limitation of the Nyquist-Shannon theorem. 

Yet another problem remains in that conventional compression methods are usually time-consuming, and beyond the processing capacity of data acquisition devices in some image processing applications, which have relatively poor hardware/software configuration and are powered by battery. How to design an appropriate sampling matrix is one of the major problems in the current compressed sensing study. In traditional compressed sensing way, random matrices are often used as the measurement matrices, such as random Gaussian matrix, random Bernoulli matrix, which meet the Restricted Isometry Property (RIP)\cite{candes2008restricted} with a large probability. 

Moreover, the traditional sampling matrix cannot actually express the structural characteristics of the signal, and another important problem of the random matrix is its difficulty to be constructed in hardware. Additionally, the difficulty in storing these matrices and certain physical constraints on the measurement process makes it challenging to realize compressed sensing in practice. When multiplying arbitrary matrices with signal vectors of high dimension, the lack of any fast matrix multiplication algorithm results in high computational cost. In this paper, we design a deep neural network (DNN) to automatically learn the required sampling matrix, which is a ternary matrix (+1, 0, -1) to reduce storage and computational costs. 

Recently, deep learning demonstrates superior performance in computer vision\cite{krizhevsky2012imagenet} \cite{girshick2015fast}. A great deal of image Compressed Sensing reconstruction methods have been developed, such as stacked denoising autoencoder (SDA)\cite{mousavi2015deep}, non-iterative reconstruction using CNN (ReconNet)\cite{kulkarni2016reconnet}, and iterative shrinkage-thresholding algorithm based network (ISTA-Net)\cite{zhang2018ista}, etc. Most traditional methods exploit some structured sparsity as an image prior and then solve a sparsity-regularized optimization problem in an iterative fashion such as basis pursuit (BP)\cite{chen1994basis}, total variation (TV)\cite{li2013efficient}, and group sparse representation (GSR)\cite{IEEEexample:zhang2014group}, etc. However, the traditional methods usually suffer highly computational complexity, and they also encounter the challenges of choosing optimal transform and tuning parameters in their solvers. The non-iterative algorithms, with deep learning method as an example, can dramatically reduce time complexity, while achieving impressive reconstruction performance, and outperform optimization-based algorithms, e.g., BP, TV, GSR, etc. 

In this paper, we propose a deep learning approach to learn an optimized ternary sampling matrix and a non-linear reconstruction mapping from measurements to the original signal. Inspired by the block compressed sensing smooth projection Landweber algorithm (BCS-SPL)\cite{fowler2011multiscale}, making up of compressed sampling, initial reconstruction and nonlinear signal reconstruction, an end-to-end deep neural network has been designed to implement the corresponding processes. Firstly, in order to design a ternary sampling matrix with low computation cost and small storage, sparse and binary constraints are applied within the proposed network architecture, by means of which the sampling matrix can be learned automatically while avoiding arduous artificial designs. Secondly, the reconstruction network consists of two parts, i.e., initial reconstruction network and a deep reconstruction network. The former uses a convolution layer, a deep separable convolution (DSC)\cite{chollet2017xception} for initial recovery, a pixel shuffle layer, which combines image block through block-based sampling to generate the initial reconstructed image. The latter further refines the initial reconstructed image to get better reconstruction quality by using dilated convolution and deep residual network. Experimental results offer a persuasive proof for the performance and the feasibility of our idea.

\section{Related Work}
Instead of sampling the entire signal, the CS theory samples a signal $x\in {{R}^{N}}$ by taking only m linear measurements, which are obtained through the following linear transformation as
$$
y=\Phi x 
\eqno{(1)}
$$
where $y$ is a $m\times 1$ measurement vector, $x$ is the original signal of size $n\times 1$ and $\Phi$ is an $m\times n$ sampling matrix. Measurement matrices commonly used in traditional CS might be Gaussian matrix\cite{IEEEexample:candes2005decoding}, Bernoulli matrix, Toeplitz matrix\cite{bryc2006spectral}, the chaotic matrix\cite{chen2018exploiting} and so on, depending on different applications. In general, recovering the original signal $x$ from its corresponding measurements $y$ is impossible, for the number of variables is much larger than the number of observations. However, if the signal $x$ is sparse in some domain $\Psi$, the CS theory shows that correct recovery of $x$ becomes possible. The most straightforward formulation of CS reconstruction can be expressed as:
$$
 \underset{x}{\mathop{\min }}\,\parallel \Psi x{{\parallel }_{p}},\ s.t.\ \ y=\Phi x
 \eqno{(2)}
$$
 where $\Psi x$ are the spare coefficients with respect to domain $\Psi$, and the subscript p is usually set to 1 or 0, characterizing the sparsity of the vector $x$. A large number of strategies have been emerged for solving this optimization problem in the literature. A special kind of them is convex optimization method, which translates the nonconvex problem into a convex one to get the approximate solution. Basis pursuit\cite{chen1994basis} is the most commonly used convex optimization method for CS reconstruction, it replaces the $l_0$ norm constraint with the $l_1$ norm one, in order to obtain the solution by solving a linear programming problem. Generally, the convex-programming methods have very high computational complexity. In essence, traditional image CS methods have high computational complexity because the requirement of iterative calculation.

To reduce the computational complexity, some greedy algorithms have been proposed for CS reconstruction. The greedy iterative algorithm aims to find each non-zero coefficient through local optimization in each iteration, including Matching Pursuit (MP)\cite{do2008sparsity}, Orthogonal Matching Pursuit (OMP)\cite{IEEEexample:tropp2007signal}, Stagewise Orthogonal Matching Pursuit (StOMP)\cite{IEEEexample:donoho2012sparse} and other various improved algorithms of OMP\cite{IEEEexample:yaghoobi2015fast}. Compared to convex-programming approaches, methods of this type exhibit relatively low computational complexity at the cost of lower reconstruction quality.

In the field of image CS, some existing works have established more complex models by exploring image priors. LI C \emph{et al.}\cite{li2013efficient} proposed the total variation (TV) regularized constraint, which is used to replace the sparsity-based one for enhancing the local smoothness. 
Zhang J \emph{et al.}\cite{IEEEexample:zhang2014group} further proposed group sparse representation (GSR) for image compressed sensing recovery by enforcing image sparsity and non-local self-similarity simultaneously. 

For sparse representation-based image CS reconstruction methods, a sampling matrix should be established first, which should be a random measurement matrix. But random matrix usually suffer a lot problems such as high computation cost, vast storage and uncertain reconstruction qualities. 

Peng H and Mi Y \emph{et al.}\cite{peng2018p} proposed the P-tensor Product CS, reducing the dimension of sampling matrix, to decrease its storage requirement, which is at the expense of more computing resources. 
Lu W \emph{et al.}\cite{IEEEexample:lu2017binary} proposed a binary method, to determine the distribution of nonzero elements of binary sampling matrix by minimizing the coherence parameters with a greedy method. Binarized matrix can obviously alleviate the storage pressure through low-bit storage, and convert multiplication into addition and subtraction through binarization (+1, -1) to mitigate the burden of calculation. But these matrices are still data independent, ignoring the characteristics of the data.
Nguyen D M \emph{et al.}\cite{nguyen2017deep} proposed a solution, utilizing deep learning, to obtain very sparse ternary projections, which can be efficiently exploited in hardware implementations for CS. However, this three-valued method introduces negative influence to the recovery performance, resulting in a poor recovery effect, that is far below that the current state-of-the-art method. 

The sensing matrix $\Phi$ usually takes the form of high dimension, especially for images. Fowler J E \emph{et al.}\cite{fowler2012block} \cite{fowler2011multiscale} proposed block-based CS (BCS) methods, which divide the input image into a couple of nonoverlapped blocks with each block sensed independently using a much smaller matrix $\Phi$. The entire image is recovered by fitting together the reconstructed blocks, followed by a complete image smoothing. However, an undesired phenomenon, i.e., block effect, emerges. Gan\cite{gan2007block} proposed block compressed sensing (BCS) for natural images, which combines block based compressed sampling and smoothed projected Landweber reconstruction.

In the past decade, a lot of deep learning-based methods have also been developed for image CS reconstruction. Mousavi \emph{et al.}\cite{mousavi2015deep} used stacked denoising autoencoder (SDA) to capture statistical dependencies and learned the representation from training data. His method improves image recovery performance as well, with the reconstruction quality comparable to traditional state-of-the-art algorithms, while significantly cut the budget of reconstruction time, which opens up a new direction for CS. 
Kulkarni \emph{et al.}\cite{kulkarni2016reconnet} proposed a deep learning-based CS reconstruction model, named ReconNet, by employment of Convolutional Neural Network (CNN) instead of SDA's fully connected network to reduce the number of parameters and computational complexity. 

By combining traditional algorithms with deep learning, and using iterative shrinkage-thresholding algorithm (ISTA), Zhang J \emph{et al.}\cite{zhang2018ista} proposed ISTA-Net to reconstruct images. He makes the use of nonlinear operations to solve the near-end sampling transformation, to reduce the computing time, and to improve the peak signal-to-noise ratio of the restored image in some specific image areas. whereas the sampling matrix of ISTA-Net is still a random matrix that is independent of the original signal. 

Sampling matrices of aforementioned methods are stored as floating-point number, which means tremendous storage and calculation pressure for hardware implementation. W Shi \emph{et al.} proposed CSNet\cite{IEEEexample:shi2019image}, which uses a deep residual convolutional neural network to realize CS, and to construct a binary sampling matrix. Although these techniques have partially improved the quality of restoration in the initial restoration stage of block compressed sensing, only image blocks are spliced and the relationship among blocks is neglected, resulting in a longer convergence time during model training.

As discussed above, CS framework consists of two parts, i.e., a sampling network and a reconstruction network, corresponding to encoding network and decoding network, respectively. The problem with the encoding side is that the sampling matrix is data-independent, and the inability to express the structural characteristics of the data affects the quality of recovery. In addition, the sampling matrix suffers from its own high computation cost as well as vast storage. With respect to the problem with the decoding network, how to reduce the block effect and further improve the quality of recovery for those who use block compressed sensing, is always concerned.

Therefore, in this paper, an end-to-end deep learning-based compressed sensing algorithm is proposed, which involves both the encoding network and the decoding network. On the encoding side, we combine binarization technology with pruning technology based on the attention mechanism to make the sampling matrix tri-valued. The decoding end consists of two networks, an initial reconstruction network using deep separable convolution and a deep reconstruction network using residual connection and dilated convolution. With such a structural design, we can mitigate the block effect and improve the quality of reconstruction.

\section{Attention-based Ternary Projection for Compressed Sensing}
In this section, the proposed end-to-end deep learning framework for compressed sensing, namely Attention-based ternary projection network (ATP-Net), is described in detail. 

The ATP-Net architecture consists of three parts, the sampling network, the initial reconstruction network and the deep reconstruction network, with block-based sampling been adopted to improve sampling efficiency and reduce hardware overhead, as illustrated by Figure.1. This method provides a general framework, which can utilize the attention mechanism\cite{zhang2019self} to obtain the ternary matrix and realize the image reconstruction. As a result, there is only a small loss in reconstruction performance compared to full-precision sampling, accompanied by a reduction of the hardware pressure, i.e., storage pressure, calculation pressure, of the sampling matrix.

\Figure[t!](topskip=0pt, botskip=0pt, midskip=0pt)[width=6.6 in]{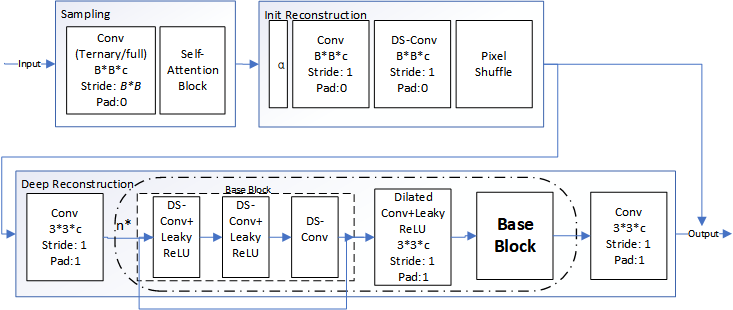}
{The proposed framework of ATP-Net.\label{fig1}}
\subsection{Sampling Network}

\subsubsection{Block-based Compressed Sampling}
Unlike traditional block-based compressed sensing (BCS), which samples input original images with non-overlapping blocks, ATP-Net employs a convolutional layer instead of a fully connected layer to spatial sampling of the input images. Mathematically, the BCS sampling can be expressed as
$$
y={{\Phi }^{big}}\otimes x
\eqno{(3)}
$$
where $\otimes$represents convolution operation and the sampling matrix $\Phi^{big}$ can be expressed as
$$
{{\Phi }^{big}}={{\left\{ \begin{matrix}
   \Phi _{1}^{sub} & \Phi _{2}^{sub} & \cdots  & \Phi _{n}^{sub}  \\
\end{matrix} \right\}}^{T}}
\eqno{(4)}
$$
where $\Phi^{big}$ denotes the weight of the whole convolutional layer, $\Phi^{sub}_{i} \in R^{bs \times bs}$ is the weight of different filters in the convolutional layer, bs means the size of each sampling block. However, we will not use activation functions and add bias units to the sampling layer to ensure that the sampling layer has only convolutional operations. As a result, the achieved linear sampling is consistent with the classic compressed sensing theory. Meanwhile, the sampling matrix can be learned adaptively during network training, which avoids the complicated manual design when constructing the sampling matrix.

 Since different sampling rates can be achieved by changing the output channel of the sampling network, which is a convolutional layer, the output channels of the sampling layer can be calculated as follows
$$
out\_chnls=blocksz*blocksz*subrate*in\_chnls
\eqno{(5)} 
$$

Here, we take a grayscale image as an example, namely image channel is 1, the sampling rate is 0.1 and block size of BCS is 32, the out channels of sampling layer will be 102. 
\subsubsection{Attention-based Ternary Projection}
Traditionally, binarization methods that seek to binarize the weights of DNN models\cite{courbariaux2015binaryconnect}, are used to address the storage and computational issues. Although the hardware pressure can be reduced by binarized weight, however, it introduces an unacceptable loss of accuracy. The ternary network reduces the storage pressure and calculation pressure of the hardware by converting the 32-bit floating-point number to the low-bit three-value (+1, 0, -1), followed by a conversion of the multiplication operation into addition and subtraction. Ternary networks seek to balance the full precision and the binary precision weight networks. In general, a ternary matrix can be achieved in two steps, namely the binarization step and the attention-based pruning step. 

The binarization step aims to convert floating point values into binarized values, i.e., mapping original weights $w_o \in R$ into binary weight $w_b \in \{+1,-1\}$. The binarization function is formulated as:
$$
 w_{b} = \begin{cases}
    +1, & if \ {w}_o \geq 0 \\
    -1, & otherwise
\end{cases} 
\label{eq6}
\eqno{(6)} 
$$

Therefore, when the data stream is forward propagated, the current weight matrix in the form of floating-point numbers is obtained, with each entry determined by formula (6). In the parameter update stage, that is, the gradient is calculated according to the corresponding binarized weights followed by the update of the floating-point number. 
In the attention-based pruning step of ATP-Net, the importance of the weight matrix will be calculated first based on the attention mechanism, and is used to prune the unimportant parameters, with the purpose to obtain a mask matrix, whose element $e \in \{ 1,0 \}$, i.e., the mask is set to 0 and 1 for the weight to be removed and kept, respectively. 

Calculation of the mask matrix is also divided into two steps. 
i) In the first step, due to the requirements of linear sampling, we establish the relationship between the attention mechanism and the weight matrix of the sampling layer through the following equation
$$
F(x)={{y}_{atten}}=\hat{W}\otimes x
\eqno{(7)}
$$
where $y_{atten}$ is the output tensor of attention layer, $\otimes$ is the convolution operation, $x$ is the input image, $\hat{W}$ is the matrix of input image after attention. Equation (7) can be solved by introducing another convolutional neural network, which optimizes the Euclidean distance between its own output tensor and that of attention layer, namely $y_{atten}$. 

ii) In the second step, the importance matrix can be calculated through the following equation
$$
Y=\hat{W}\ominus W
\eqno{(8)}
$$
where $\ominus$ is the element-wise subtraction between matrices, $W$ is the weight matrix of sampling layer and the elements in $Y$ represent the importance of their corresponding weights. We will prune elements with smaller values $y_i \in Y$ according to the sparsity rate. Finally, the pruning is realized by 
$$
{{W}_{t}}={{W}_{b}}*Mask
\eqno{(9)}
$$
where $*$ is the Hadamard product operation, $W_b$ is ternary matrix, $W_b$ is binary matrix and Mask is the pruning mask based on attention method. 

\subsubsection{Scaling Factor Alpha}
In order to compensate for the loss of precision caused by ternary matrix, we use an adaptive scaling factor, alpha, to enhance the expressive ability of ternary matrix. This scaling factor learns itself adaptively during model training.

\subsection{Reconstruction Network}
\subsubsection{Depth Separable Convolution}
In order to fully express the relationship between the various feature maps, the initialization recovery module uses a depth separable convolution layer to replace the standard convolution layer in the initial reconstruction module. The depth separable convolution\cite{chollet2017xception} consists of two steps, deep convolution and point-wise convolution. Different convolution kernels can be used for each input feature map, which can express the relationship within the feature map and that between the feature maps of each layer.

By decoupling the channel dimension of the feature map and the interior relationship of the feature map, it is possible to obtain not only the internal local features in the channel, but also the channel features between different feature maps. The realization of the depth separable convolution emphasizes on extraction of the high-level features of the input tensor by combining deep convolution and point-wise convolution through two steps.

For deep convolution, the schematic diagram of the convolution mode is shown in Figure2. Each filter in the deep convolution is responsible for a single channel. Unlike unconventional convolutions operate in that each channel is processed by only one filter, conventional ones convolve the input and then perform the adding processing. The number of feature maps generated by this process is the same as that of input channels, and that of channels which will not change.
\Figure[t!](topskip=0pt, botskip=0pt, midskip=0pt){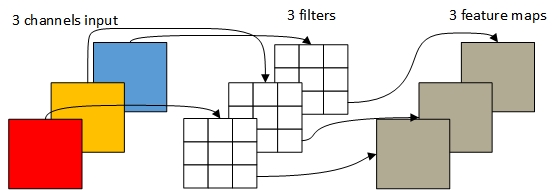}
{Diagram of channel-by-channel convolution diagram.\label{fig2}}

For the separated point-by-point convolution, a convolution kernel of size 1 is introduced to perform conventional convolution operations upon the input, pixel by pixel. As a result, the convolution operation of point-by-point convolution is capable of learning the weights of the input tensor in the channel direction, and generate a new feature map for output. In general, dimension of the input tensor will not change, both horizontally and vertically, through the module, using depth separable convolution without adding boundary filling.

\subsubsection{Dilated Convolution}
Dilated convolutions\cite{yu2015multi} and ordinary ones differ in the convolution kernel. The latter uses compact kernel, while the former can control the internal interval of its convolution kernel through the expansion rate, whose mathematical description can be shown as follows:
$$
{{y}_{i}}=\sum\limits_{k=1}^{K}{{{x}_{i+k\times r}}}{{w}_{k}}
\eqno{(10)}
$$
where $x$ is the input one-dimensional signal, the size of the convolution kernel used for dilation convolution is $k$, and the expansion rate is denoted by $r$.

Figure 3 is a principled diagram of the aforementioned two convolutions, which demonstrates their difference of the kernel structures, revealed by -a and -b, respectively. It can be seen from the figure that dilated convolution is a special case of ordinary convolutions, by padding 0 to expand convolution kernel. Consequently, the dilated convolution can expand the receptive field arbitrarily without introducing additional parameters.

\Figure[t!](topskip=0pt, botskip=0pt, midskip=0pt){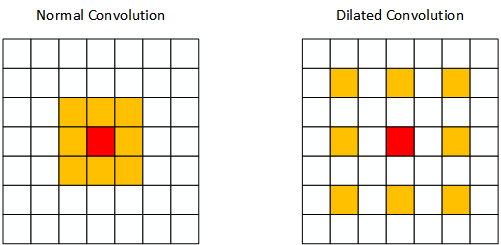}
{Diagram of dilated convolution.
\label{fig3}}

Dilated convolutions also play a very beneficial role in removing the block effect of block compressed sensing. Because the block effect is caused by the loss of context information as well as the resulted edge effect, it can be reduced by adding dilated convolution. Simultaneously, the reconstruction quality can be improved.

However, continuous use of dilated convolutions will trigger a chessboard problem. The reason is that in the feature map of the output of a dilated convolution, neighborhood pixels of a specified pixel can be obtained by convolution from different independent sets. For the problem of continuous cascading of dilated convolutional layers, a combination of different expansion rates with a common divisor greater than 1 is conducted. Although the method of Hybrid Dilated Convolution\cite{wang2018understanding} can be used, an excessive expansion rate requires a large boundary filling, leading to unnecessary operations, which is the reason that interspersed basic blocks are used to avoid cascading and chessboard problems.

\subsubsection{Reconstruction Network Architecture}
The initial reconstruction part, in which a depth separable convolution is used as the basic structure, is used to obtain a simple reconstruction of the image from the compressed vector. The reason why depth separable convolution is used instead of ordinary convolution is that an ordinary convolution can merely express the features within each layer of feature maps and ignores the data relevance among feature maps. However, the input data of the initial reconstruction module is measurement of block compressed sampling. There is a clear physical meaning between these layers, i.e., the spatial association between different sampling blocks. Therefore, the features between layers can be learned to estimate the correlation between sampling blocks, by using depth separable convolutions, learning the features in layers, mitigating the block edge effect caused by the block sampling, and also reducing the number of parameters of the model.

The deep reconstruction part uses a stack of basic blocks based on channel splicing, 2D convolution, and residual connections, which is used to further optimize quality of the image reconstruction. The residual connections within the blocks and the residual connections between blocks are used to avoid network training problem induced by excessive network layers. At the same time, we add a dilated convolutional layer before one or several basic blocks. By using the dilated convolutional layer, the number of intervals between the convolution kernels can be increased by setting a certain dilation rate to expand the receptive field. Compared with traditional down sampling methods, such as adding a pooling layer, etc., dilated convolution can increase the receptive field without degrading the resolution and is able to capture more contextual information. Adding a dilated convolution layer between multiple basic blocks is beneficial to circumvent the problem caused by uninterrupted use of dilated convolution.

The output of initial reconstruction, $\hat{X}$ , is fed as the input to the deep reconstruction part, on which the feature extraction is performed as follows:
$$
 {{X}^{c}}=F({{W}_{c}}\cdot \hat{X}+{{b}_{c}}) 
\eqno{(11)}
$$
where $F(\centerdot )$ represents the activation function and LeakyReLU is employed, ${{b}_{c}}$ represents the bias of the layer.

After been processed by the first convolutional layer, the signal will go through multiple basic blocks and dilated convolution one by one for feature learning, which can be formally expressed as

$$
 X_{i}^{base}=F(g(X_{i}^{c})+X_{i-1}^{base})
\eqno{(12)}
$$
$$
{{X}^{d}}=F({{W}_{d}}\cdot X_{i}^{c}+{{b}_{d}}) 
\eqno{(13)}
$$
where $X_{i}^{base}$ represents the output result of the i-th basic block, $g(\centerdot )$ represents the basic block operation, and ${{X}^{d}}$ represents the output of the dilated convolution.

Composition of the basic block in the deep reconstruction module is inspired by the ESRGAN\cite{wang2018esrgan} \cite{rakotonirina2020esrgan+} series networks in the field of image super-resolution, combined with the idea of Dense Block in the internal basic structure of the ESRGAN series of networks, named RRDB (Residual in Residual Dense Block). The input and output of each layer are concatenated to form the input of its successive layer, as illustrated by Figure 4, which demonstrates the detailed structure diagram of the base block.

\Figure[t!](topskip=0pt, botskip=0pt, midskip=0pt){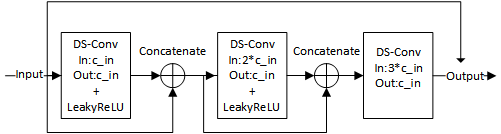}
{Base Block structure diagram.\label{fig4}}

In Figure 4, \emph{DS-Conv} represents the depth separable convolutional layer, as that in Figure4. \emph{In} and \emph{Out} represent the number of input channels and output ones of the convolutional layer, respectively, and \emph{Concatenate} represents the channel concatenation of two tensors. The process of basic block, i.e., from input to output, can be expressed as

$$
	{{y}_{1}}=F(W_{d}^{1}\cdot x+b_{d}^{1})
\eqno{(14)}
$$
$$
{{y}_{2}}=F(W_{d}^{2}\cdot ({{y}_{1}}\oplus x)+b_{d}^{2})
\eqno{(15)}
$$
$$
{{y}_{3}}=F(W_{d}^{3}\cdot ({{y}_{2}}\oplus {{y}_{1}}\oplus x)+b_{d}^{3})
\eqno{(16)}
$$
$$
g(x)={{y}_{3}}+\beta x
\eqno{(17)}
$$
where $x$ is the input, and defined as a quadruple, {batch, channel, height, width}.   and  represent the weight matrix and bias of the \emph{i-th} hidden layer, respectively.  represents the $Concatenate$ operation in Figure 4. The number of input channels of the convolutional layer is increased by \emph{c\_in} for each deepening layer, where \emph{c\_in} is the number of channels of $x$. Finally, and times $x$ are subjected to residual calculation. 
\section{Experimental Results}
\subsection{Dataset}
We use the BSDS500 database\cite{martin2001database}, which consists of 500 images, and LIVE1 database\cite{sheikh2005live} \cite{IEEEexample:sheikh2006statistical} \cite{IEEEexample:wang2004image} for training and validation, respectively. In order to enhance the diversity of data and improve the performance of the model, TorchVision\cite{marcel2010torchvision} are utilized to preprocess data and achieve data enhancement. The input training image is randomly cropped, with image size set to 96×96, and the horizontal and vertical directions are both randomly flipped with the probability of 50\% . Finally, the grayscale processing is performed to obtain the final input data that will be fed to the ATP-Net. Meanwhile, we select Set11\cite{kulkarni2016reconnet} to evaluate the proposed methods, which consists of 11 grayscale images.

\subsection{Training details of ATP-Net model}
In the experiment, the ATP-Net model is built and trained based on PyTorch, with hyperparameters set as follows: the block size at the sampling part is 32, the initial learning rate is set to 1×10-4, the batch size is 32, and the learning rate is reduced to 10\% of the original learning rate every 100 rounds. It should be noted that the length and width of the input image must be divisible by 32, since the block size of block sampling is set to 32 pixels. Otherwise, the input image has to be cropped. In case the length or width of the input image is less than 32 pixels, no output can be obtained.

\subsection{Comparison with state of the art}
In this paper, performance of ATP-Net will be evaluated in terms of both objective and subjective assessment, i.e., peak signal-to-noise ratio (PSNR) and visual evaluation, with TVAL3 algorithm, D-AMP algorithm, ReconNet and DR2-Net being selected as references. As for the Set11\cite{kulkarni2016reconnet} dataset, the measurement rates (MR) of 0.25, 0.1, 0.04, 0.01 are used. The Set11\cite{kulkarni2016reconnet} dataset is mainly composed of natural images, including classic images used in the field of image measurement, such as Lena, etc., as well as animated images \emph{Flinstones}, to test the model's capability of reconstruction in unnatural images.

Table 1 lists the PSNR performance of different algorithms at various sampling rates. It can be seen from Table 1 that the improvement of ATP-Net is obvious at a lower sampling rate, and it tends to be less and less obvious as the sampling rate increases, which implies that ATP-Net behaves better at lower sampling rate.

\begin{table}
\caption{PSNR of Set11\cite{kulkarni2016reconnet}}
\begin{tabular}{llllll}
Image       & Mothed   & MR=0.25        & MR=0.10        & MR=0.04        & MR=0.01        \\
Barbara     & TVAL3    & 24.19          & 21.88          & 18.98          & 11.94          \\
            & D-AMP    & 25.08          & 21.23          & 16.37          & 5.48           \\
            & ReconNet & 23.25          & 21.89          & 20.38          & 18.61          \\
            & DR2-Net  & 25.77          & 22.69          & 20.7           & 18.65          \\
            & ATP-Net  & \textbf{26.81} & \textbf{24.18} & \textbf{22.94} & \textbf{21.36} \\
Fingerprint & TVAL3    & 22.7           & 18.69          & 16.04          & 10.35          \\
            & D-AMP    & 25.17          & 17.15          & 13.82          & 4.66           \\
            & ReconNet & 25.57          & 20.75          & 16.91          & 14.82          \\
            & DR2-Net  & 27.65          & 22.03          & 17.4           & 14.73          \\
            & ATP-Net  & \textbf{28.91} & \textbf{25.42} & \textbf{19.55} & \textbf{16.1}  \\
Flinstones  & TVAL3    & 24.05          & 18.88          & 14.88          & 9.75           \\
            & D-AMP    & 25.02          & 16.94          & 12.93          & 4.33           \\
            & ReconNet & 22.45          & 18.92          & 16.3           & 13.96          \\
            & DR2-Net  & 26.19          & 21.09          & 16.93          & 14.01          \\
            & ATP-Net  & \textbf{27.56} & \textbf{24.42} & \textbf{19.36} & \textbf{16.35} \\
Lena        & TVAL3    & 28.67          & 24.16          & 19.46          & 11.87          \\
            & D-AMP    & 28             & 22.51          & 16.52          & 5.73           \\
            & ReconNet & 26.54          & 23.83          & 21.28          & 17.87          \\
            & DR2-Net  & 29.42          & 25.39          & 22.13          & 19.97          \\
            & ATP-Net  & \textbf{31.05} & \textbf{28.53} & \textbf{24.97} & \textbf{21.7}  \\
Monarch     & TVAL3    & 27.77          & 21.16          & 16.73          & 11.09          \\
            & D-AMP    & 26.39          & 19             & 14.57          & 6.2            \\
            & ReconNet & 24.31          & 21.1           & 18.19          & 15.39          \\
            & DR2-Net  & 27.95          & 23.1           & 18.93          & 15.33          \\
            & ATP-Net  & \textbf{30.17} & \textbf{27.55} & \textbf{22.32} & \textbf{20.47} \\
Parrots     & TVAL3    & 27.17          & 23.13          & 18.88          & 11.44          \\
            & D-AMP    & 26.86          & 21.64          & 15.78          & 5.09           \\
            & ReconNet & 25.59          & 22.63          & 20.27          & 17.63          \\
            & DR2-Net  & 28.73          & 23.94          & 21.16          & 18.01          \\
            & ATP-Net  & \textbf{30.52} & \textbf{27.63} & \textbf{23.99} & \textbf{21.85} \\
Boats       & TVAL3    & 28.81          & 23.86          & 19.2           & 11.86          \\
            & D-AMP    & 29.26          & 21.9           & 16.01          & 5.34           \\
            & ReconNet & 27.3           & 24.15          & 21.36          & 18.49          \\
            & DR2-Net  & 30.09          & 25.58          & 22.11          & 18.67          \\
            & ATP-Net  & \textbf{31.47} & \textbf{28.85} & \textbf{24.66} & \textbf{21.68} \\
Cameraman   & TVAL3    & 25.69          & 21.91          & 18.3           & 11.97          \\
            & D-AMP    & 24.41          & 20.35          & 15.11          & 5.64           \\
            & ReconNet & 23.15          & 21.28          & 19.26          & 17.11          \\
            & DR2-Net  & 25.62          & 22.46          & 19.84          & 17.08          \\
            & ATP-Net  & \textbf{28.05} & \textbf{25.35} & \textbf{22.19} & \textbf{20.03} \\
Foreman     & TVAL3    & 35.42          & 28.69          & 20.63          & 10.97          \\
            & D-AMP    & 35.45          & 25.51 n         & 16.27          & 3.84           \\
            & ReconNet & 29.47          & 27.09          & 23.72          & 20.04          \\
            & DR2-Net  & 33.53          & 29.2           & 25.34          & 20.59          \\
            & ATP-Net  & \textbf{35.28} & \textbf{33.71} & \textbf{30.07} & \textbf{25.68} \\
House       & TVAL3    & 32.08          & 26.29          & 20.94          & 11.86          \\
            & D-AMP    & 33.64          & 24.55          & 16.91          & 5              \\
            & ReconNet & 28.46          & 26.69          & 22.58          & 19.31          \\
            & DR2-Net  & 31.83          & 27.53          & 23.92          & 19.61          \\
            & ATP-Net  & \textbf{34.08} & \textbf{31.86} & \textbf{27.41} & \textbf{23.17} \\
Peppers     & TVAL3    & 29.62          & 22.64          & 18.21          & 11.35          \\
            & D-AMP    & 29.84          & 21.39          & 16.13          & 5.79           \\
            & ReconNet & 24.77          & 22.15          & 19.56          & 16.82          \\
            & DR2-Net  & 28.49          & 23.73          & 20.32          & 16.9           \\
            & ATP-Net  & \textbf{31.04} & \textbf{27.56} & \textbf{23.35} & \textbf{20.47} \\
Mean PSNR   & TVAL3    & 27.84          & 22.84          & 18.39          & 11.31          \\
            & D-AMP    & 28.17          & 21.14          & 15.49          & 5.19           \\
            & ReconNet & 25.54          & 22.68          & 19.99          & 17.27          \\
            & DR2-Net  & 28.66          & 24.32          & 20.8           & 17.44          \\
            & ATP-Net  & \textbf{30.4}  & \textbf{27.73} & \textbf{23.71} & \textbf{20.55}
\end{tabular}
\end{table}

Figure 5 compares the reconstruction effect of the ATP-Net model over the Set11 dataset, which demonstrates the original images in (a), and images generated by ATP-Net with sampling rate set to 0.25 in (b). It can be seen that 11 images in Set11 dataset as whole have a fairly satisfactory reconstruction effect, with a few being polluted by a certain amount of noise due to the loss caused by the ternarization. For an animated image of \emph{FLINSTONES}, the reconstruction effect is relatively ordinary, since this type of images differ from commonly natural images, which are usually used as the training dataset. Table 1 shows the corresponding PSNR index values.

\Figure[t!](topskip=0pt, botskip=0pt, midskip=0pt)[width=6.6 in]{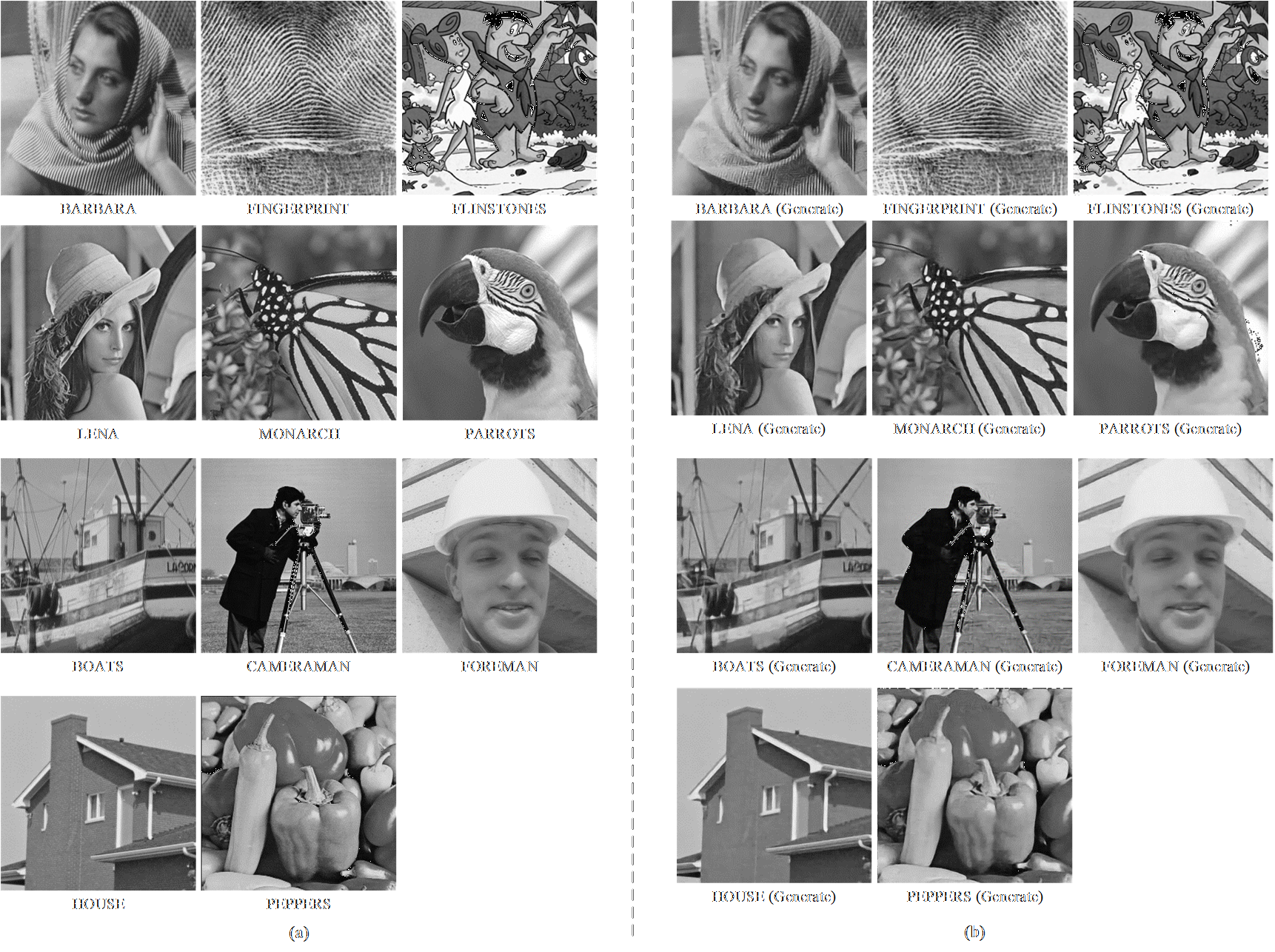}
{Visual quality comparisons of ATP-Net reconstruction on Set11 \cite{kulkarni2016reconnet} where measurement rate = 0.25.\label{fig5}}

\section{Conclusion}
In this paper, we propose a ternary sampling matrix method based on attention mechanism and implements a compressed sensing algorithm, named ATP-Net model, for image reconstruction. The model ternarizes the sampling matrix to alleviate the problems of large storage requirement and low computational efficiency of the sampling matrix in compressed sensing, on the basis of perceiving the importance of the weights, which based on the binarization of the sampling matrix, through the attention mechanism. Experimental results over the Set11 dataset illustrate that ATP-Net out-performances the state-of-the-art algorithms, e.g., TVAL3, D-AMP, ReconNet and DR2-Net, and so forth, either in terms of PSNR, or by means of subjective comparison.

\bibliographystyle{IEEEtran}
\bibliography{IEEEabrv,IEEEexample}

\begin{IEEEbiography}[{\includegraphics[width=1in,height=1.25in,clip,keepaspectratio]{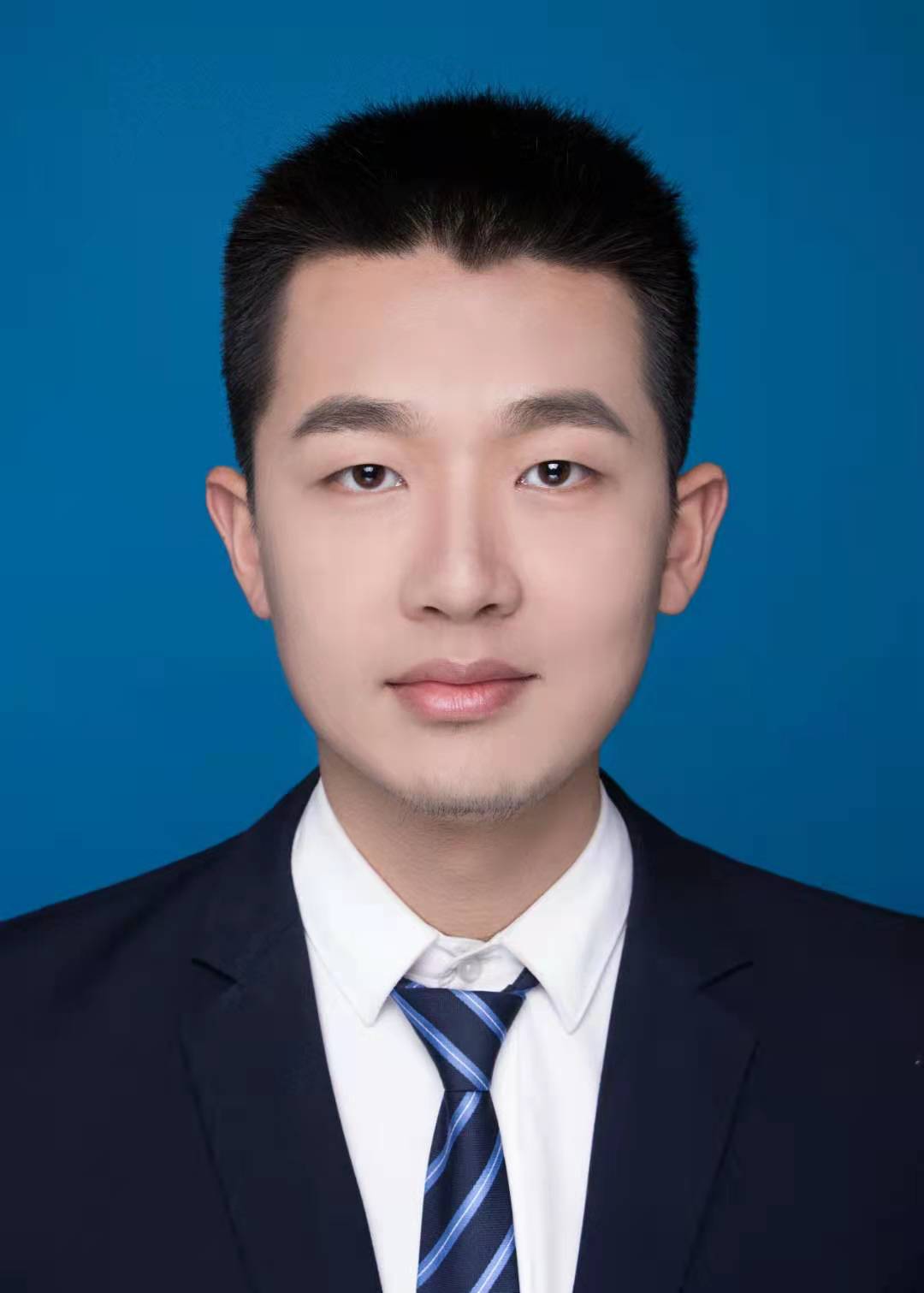}}]{First A. Guanxiong Nie}  
    received an M.S. degree in Cyberspace Security from Beijing University of Posts and Telecommunications, China in 2021. His research interests include compressed sensing, statistical learning theory and green model. Mr. Nie is a student member of the Chinese Association for Artificial Intelligence (CAAI).
\end{IEEEbiography}

\begin{IEEEbiography}[{\includegraphics[width=1in,height=1.25in,clip,keepaspectratio]{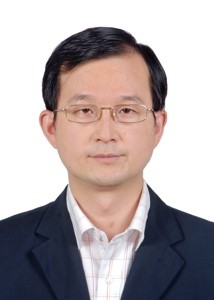}}]{SECOND B. Yajian Zhou} 
    was born in China in 1971. He obtained his Ph.D. degree in Communications Engineering 
    from Xidian University at Xi’an in 2003. He received his M.S. degree in 1996 and B.S. 
    degree in 1993, both from Beihang University. He is currently an associate professor in 
    the School of Cyberspace Security of Beijing University of Posts and Telecommunications 
    (BUPT). His main research interests include security of wireless networks, natural 
    language processing, privacy protection-based data mining, etc. He is now a main member 
    of a project of the National Key Research and Development Program of China. He was once 
    the project leaders of the National 863 High-Tech Research and Development Plan of China, 
    the National Natural Science Foundation of China and a project from Beijing Municipal 
    Natural Science Foundation. He has published approximately 50 technical papers and 6 
    books.
\end{IEEEbiography}
\EOD

\end{document}